\documentclass[aps,pre,twocolumn,groupedaddress]{revtex4-2}

\usepackage{amsmath}
\usepackage{graphicx}

\begin{document}


\title{Nonequilibrium steady states in bead-spring models: \\Entropy production and probability distributions}


\author{Jetin E Thomas}
\email[]{jetinthomas@iisermohali.ac.in}
\affiliation{Department of Physical Sciences, Indian Institute of Science Education and Research Mohali, \\ Knowledge city, Sector 81, Manauli, PO, Sahibzada Ajit Singh Nagar, Punjab 140306, India}
\author{Ramandeep S. Johal}
\email[]{rsjohal@iisermohali.ac.in}
\affiliation{Department of Physical Sciences, Indian Institute of Science Education and Research Mohali, \\ Knowledge city, Sector 81, Manauli, PO, Sahibzada Ajit Singh Nagar, Punjab 140306, India}


\date{\today}

\begin{abstract}
We study non-equilibrium models comprising of beads connected by springs. The
system is coupled to two thermal baths kept at different temperatures. We derive the steady state
probability distributions of positions of the bead for the 
one-bead system in the underdamped case. We employ the recently proposed technique of an effective temperature, along with  numerical simulations to solve the Langevin equations and obtain their corresponding probability distributions. It is observed that the marginal probability distributions in the position are independent of mass.  We also obtain theoretically and numerically the rate of entropy production for the one-bead system. The probability distribution of the positions in the two-beads system are obtained theoretically and numerically, both in the underdamped and overdamped case. Lastly, we discuss the notion of ergodicity and have tested the convergence of the 
time-averaging and the ensemble-averaging protocols.  
\end{abstract}


\maketitle

\section{\label{Introduction} Introduction}
The bead-spring model with two heat baths has a more recent but significant place in the history of non-equilibrium statistical mechanics, emerging primarily in the late 20th and early 21st century as a minimal model for nonequilibrium steady states (NESS). This model sits at the interface of equilibrium and nonequilibrium physics as it is simple enough to solve exactly,  yet rich enough to exhibit irreversibility, circulating currents, and heat flows.
The model originated in polymer physics, e.g., the Rouse model  \cite{rouse1953theory}, and was later extended to non-equilibrium settings by introducing multiple heat baths at different temperatures. In the 1990s and early 2000s, physicists sought simple, analytically tractable models to understand heat conduction at small scales, entropy production, fluctuation theorems, and steady-state thermodynamics. This model became a workhorse model for capturing key non-equilibrium features while allowing exact calculations \cite{lebowitz1999gallavotti}. Dhar studied 
bead-spring chains under thermal gradients \cite{dhar2008heat}. Van den Broek, Kawai, and Esposito from the years starting from 2004 to 2014 used single-particle models with two heat baths to derive the rate of entropy production, heat currents, and Lyapunov-based covariance matrices \cite{van2004microscopic}.  


The minimal model comprises of beads connected by springs and  coupled to two heat baths kept at different temperatures. 
The system parameters for the bead-spring models constitute the stiffness constants of the different springs, the temperatures of the heat baths, the masses of the beads, and the coefficients of frictional drag \cite{van2004microscopic, tome2006entropy, verley2014unlikely}. It is an analytically tractable model whose joint probability distribution of the bead positions and velocities could be determined by solving the continuous Lyapunov equation \cite{herzel1991risken, gardiner1985handbook}. The Lyapunov equation helps us in determining the covariance matrix \cite{herzel1991risken, gardiner1985handbook}. This covariance matrix is the primary component needed in determining the probability distributions \cite{van2004microscopic, seifert2012stochastic, barato2015thermodynamic}. Recently, Wu and Wang described the non-equilibrium equation of state for 
an $N$-harmonic chain maintained in an NESS,  by defining equilibrium-like thermodynamic quantities \cite{wu2022nonequilibrium}. Extending this work, 
Tu \cite{tu2025weighted} studied the notion of an effective temperature  for such a bead-spring model where the joint probability distribution could be written in terms of a Boltzmann weight using an effective hamiltonian and an effective temperature. 

In this paper, we compute the probability distributions for the one- and two-beads systems from different approaches based on i) an effective temperature, and ii) solving the Langevin equations for these models  numerically.  We keep track of the nonequilibrium effects on the surroundings by measuring the rate of entropy production \cite{seifert2012stochastic, van2015ensemble, tome2015stochastic}. A lot of work on the overdamped case \cite{li2019quantifying, tome2006entropy} exists in literature  where the mass of the bead can be neglected. We perform our study in the underdamped case, with a finite mass, and then take the zero-mass limit to compare the results with the overdamped case. We find  that the probability distribution for the bead position  in the one-bead set up is independent of  mass. Thereby, the probability distributions in the overdamped limit come exactly equal to the distributions obtained in the underdamped case. However, this is not the case for a two-beads set up where the distributions do depend on mass. Also for the latter set up, the approach of an effective temperature  \cite{tu2025weighted} for the distributions comes exactly to those obtained from the numerical simulations.

Further, we discuss the convergence of results obtained from time-averaging and ensemble-averaging. We find that the rate of convergence to NESS is faster for algorithms based on the former as compared to the latter. 
We have also considered the equivalence of both the protocols by comparing the rate of entropy production and the marginal probability distributions of positions and velocities.

The plan of the paper is as follows.
We first describe the steady-state probability distributions of positions for the one-bead system in Sec.  \ref{SSPD_OneBead}. We use different approaches to derive this probability distribution. These approaches are described in subsection  \ref{CPDA_OneBead} of Sec.  \ref{SSPD_OneBead}. We also numerically calculate the rate of entropy production and compare it with that obtained from theory in subsection  \ref{Ent_OneBead}. Next, we focus on the probability distribution of the beads in the two-beads systems in Sec.  \ref{SSPD_TwoBead}. In subsection  \ref{CPDA_TwoBead}, we describe the various approaches used for the calculation of probability distributions and describe the case arising from the approach using effective temperature. We also derive the probability distributions for the overdamped limit and compare the probability distributions for beads having mass where the mass is taken to be really small. In the subsection \ref{Ent_TwoBead}, we numerically compute the rate of entropy production for a two-beads system and compare it with the 
one-bead system. Finally, in Sec. \ref{ergodicity} we  
discuss  the convergence of results implying the equivalence of time averaging and ensemble averaging protocol.
%
\section{\label{SSPD_OneBead} Steady-state probability distribution for one-bead system}
\begin{figure}
\includegraphics[scale=0.4]{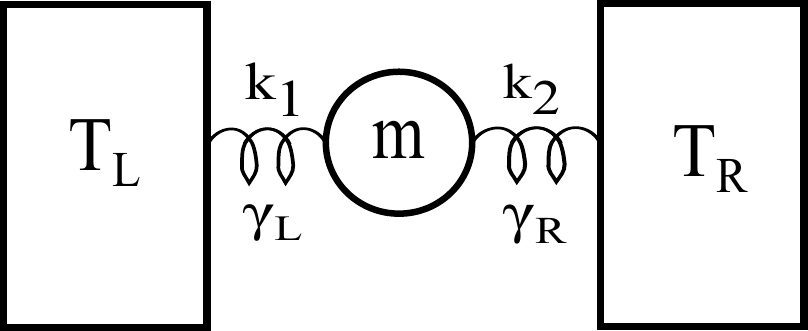}
\caption{A bead with mass $m$ simultaneously coupled to two baths at temperatures $T_{L}$ and $T_{R}$, with springs having spring constants $k_{1}$ and $k_{2}$. The frictional drag coefficients are given by $\gamma_{L}$ and $\gamma_{R}$.}
\label{onebeadsetup}
\end{figure}
In this section, 
we calculate the steady-state probability distributions 
based on the concept of effective temperature. We develop the machinery using the covariance matrix ($\boldsymbol{\sigma}$), modified covariance matrix ($\tilde{\boldsymbol{\sigma}}$), and residual matrix ($\tilde{\boldsymbol{\sigma}}^{r}$) to come up with an effective temperature ($T_{e}$) from the Langevin equations similar in lines to Tu's work in \cite{tu2025weighted}. The model consists of 
a bead simultaneously coupled to two  thermal reservoirs with temperatures $T_{L} > T_{R}$ by  springs 
with spring constant $k_{1}$ and $k_{2}$, and 
frictional coefficients $\gamma_L$ and $\gamma_R$
respectively 
as shown in Fig. (\ref{onebeadsetup}). 
The Langevin set of equations for this system \cite{maes2014nonequilibrium, van2015ensemble, tu2025weighted} is given by
\begin{align}
\frac{dx_{1}}{dt} = & \;  v_{1}, \label{pos_Langevin_1Bead} \\ \nonumber
\frac{dv_{1}}{dt} = & -\frac{k_{1}x_{1}}{m}-\frac{k_{2}x_{1}}{m}\\
&+\left[-\frac{\gamma_{L}v_{1}}{m}+\xi_{L}(t)\right]+\left[-\frac{\gamma_{R}v_{1}}{m}+\xi_{R}(t)\right] , \label{vel_Langevin_1Bead}
\end{align}
where $x_{1}$ is the displacement of the particle from the equilibrium position, and $v_{1}$ is the velocity of the particle. $\xi_{L}(t)$ and $\xi_{R}(t)$ are white noises due to each reservoir, which satisfy $\langle \xi_{\alpha}(t) \rangle = 0$ and $\langle \xi_{\alpha}(t)\xi_{\alpha}(t') \rangle = (2{\gamma_{\alpha}k_{B}T_{\alpha}}/{m^{2}})\delta(t-t')$ with 
$\alpha=L,R$ and $m$ as the mass of the bead. We have set $k_{B} = 1$ in the rest of the paper. The physical units for the degrees of freedom and system parameters could be appropriately chosen to fit the description of the langevin model (eqs. (\ref{pos_Langevin_1Bead}) and (\ref{vel_Langevin_1Bead})).

Now, the Langevin equations  (\ref{pos_Langevin_1Bead}) and (\ref{vel_Langevin_1Bead}) represent the Ornstein-Uhlenbeck process where the joint probability distribution for $x_{1}$ and $v_{1}$ is given as
\begin{equation}
P( \boldsymbol{z} = [x_{1},v_{1}]^{T}) = 
\frac{\exp ({-{\boldsymbol{z}^{T}\boldsymbol{\sigma}^{-1}\boldsymbol{z}}/{2}})}{\sqrt{{\rm Det}[ 2\pi \boldsymbol{\sigma}]}},
\label{Prob_interms_cov_matrix}
\end{equation}
where $\boldsymbol{\sigma}^{-1}$ is the inverse of
the covariance matrix ($\boldsymbol{\sigma}^{}$) \cite{van2004microscopic, seifert2012stochastic, barato2015thermodynamic}. 
In Appendix A, we have described formulation of the effective temperature ($T_{e}$) from the Langevin equations for the one-bead system, which is used to write the joint probability distribution in the form: 
$P(\boldsymbol{z}) = \exp({-[H+\Delta H]/T_{e}})/\mathcal{Z}$ where $H=\frac{1}{2}mv_{1}^{2}+\frac{1}{2}k x_{1}^{2}$ and in the present case, $\Delta H=0$.
 $k =k_{1}+k_{2}$ is the effective spring constant. 

\subsection{\label{CPDA_OneBead} Comparison of probability distributions from different approaches}
Based on eqs. (\ref{Modified_Covariance_Matrix_1D}) and
(\ref{KTilde_matrix}) in Appendix A, we find the covariance matrix ($\boldsymbol{\sigma}$) from the modified covariance matrix ($\boldsymbol{\tilde{\sigma}}$) as
\begin{equation}
\boldsymbol{\sigma} = \tilde{\boldsymbol{\sigma}}\tilde{\boldsymbol{K}}^{-1} = T_e\begin{bmatrix}
1/k & 0 \\
0 &  1/m
\end{bmatrix},
\label{CovMatrix_OneBead}
\end{equation}
where $T_e$ is an effective temperature for the one-bead case, given by Eq. (\ref{Teff_OneBead}).
The joint probability distribution 
is then evaluated as
\begin{eqnarray}
P(\boldsymbol{z}) = \frac{\sqrt{mk}}{2\pi T_{e}}e^{-(mv_{1}^{2}+kx_{1}^{2})/2T_{e}}. 
\label{Prob_Cov_Matrix}
\end{eqnarray}
%
%
When the mass is tiny, we can take the overdamped limit of $m \rightarrow 0$, and the Langevin equation in eq. (\ref{vel_Langevin_1Bead}) becomes

\begin{align}
\frac{dx_{1}}{dt} &= -\frac{k}{\gamma_{L}+\gamma_{R}}x_{1}+\frac{\xi_{L}(t)+\xi_{R}(t)}{\gamma_{L}+\gamma_{R}}, 
\nonumber \\
& = -\frac{k}{\gamma_{e}}x_{1}+ \sqrt{2\frac{D}{\gamma_{e}^{2}}}\xi(t).
\label{vel_Langevin_od}
\end{align}
Here, 
$D=\gamma_{L}T_{L}+\gamma_{R}T_{R}$ and $\gamma_{e} = \gamma_{L}+\gamma_{R}$.
The sum of two noises is effectively a single 
noise term ($\xi(t)$) with the correlation $\langle \xi(t)\xi(t') \rangle \propto \delta(t-t')$.  We can write the Fokker-Planck equation for the probability distribution ($P_{od}$) for the Langevin equation  (\ref{vel_Langevin_od}) as \cite{van1992stochastic, risken1996Fokker, van2006brownian, seifert2008stochastic}
\begin{equation}
\frac{\partial P_{od}}{\partial t}= \frac{k}{\gamma_{e}}\frac{\partial (x_{1}P_{od})}{\partial x_{1}}+ \frac{D}{\gamma_{e}^{2}}\frac{\partial^{2} P_{od}}{\partial x_{1}^{2}}.
\label{Fokker_Planck_od}
\end{equation}
In the steady state, $\partial P_{od}/\partial t = 0$, and assuming the solution in the form 
$P_{od}(x_{1}) = B \exp ({-\zeta x_{1}^{2}})$, we get, from  eq. (\ref{Fokker_Planck_od}), the condition: 
%
$\left({k}{\gamma_{e}}- 2\zeta D \right)\left(1-2\zeta x_{1}\right) = 0$.
Since, this equation should be satisfied for any position ($x_{1}$) of the bead, it implies 
$\zeta = {k\gamma_{e}}/{2D} = {k}/{2 T_{e}}$. The normalizing constant $B$ comes as $\sqrt{k/2\pi T_e }$. Thus, the overdamped steady-state probability distribution is given by
\begin{equation}
P_{od}(x_{1}) = \frac{1}{\sqrt{2\pi T_{e}/k}}
\exp({-{kx_{1}^{2}}/{2 T_{e}}}).
\label{probod}
\end{equation}
Note that $P_{od}(x_{1})$ is also equal to the marginal probability distribution as defined by $P(x_{1}) = \int_{-\infty}^{\infty}P(\boldsymbol{z})dv_{1}$, where $P(\boldsymbol{z})$ is given by eq. (\ref{Prob_Cov_Matrix}). 


We also  solve the Langevin equation in eqs. (\ref{pos_Langevin_1Bead}) and  (\ref{vel_Langevin_1Bead})
numerically in order to obtain the marginal distribution 
$P(x_1)$. 
\begin{figure}
\includegraphics[scale=0.32]{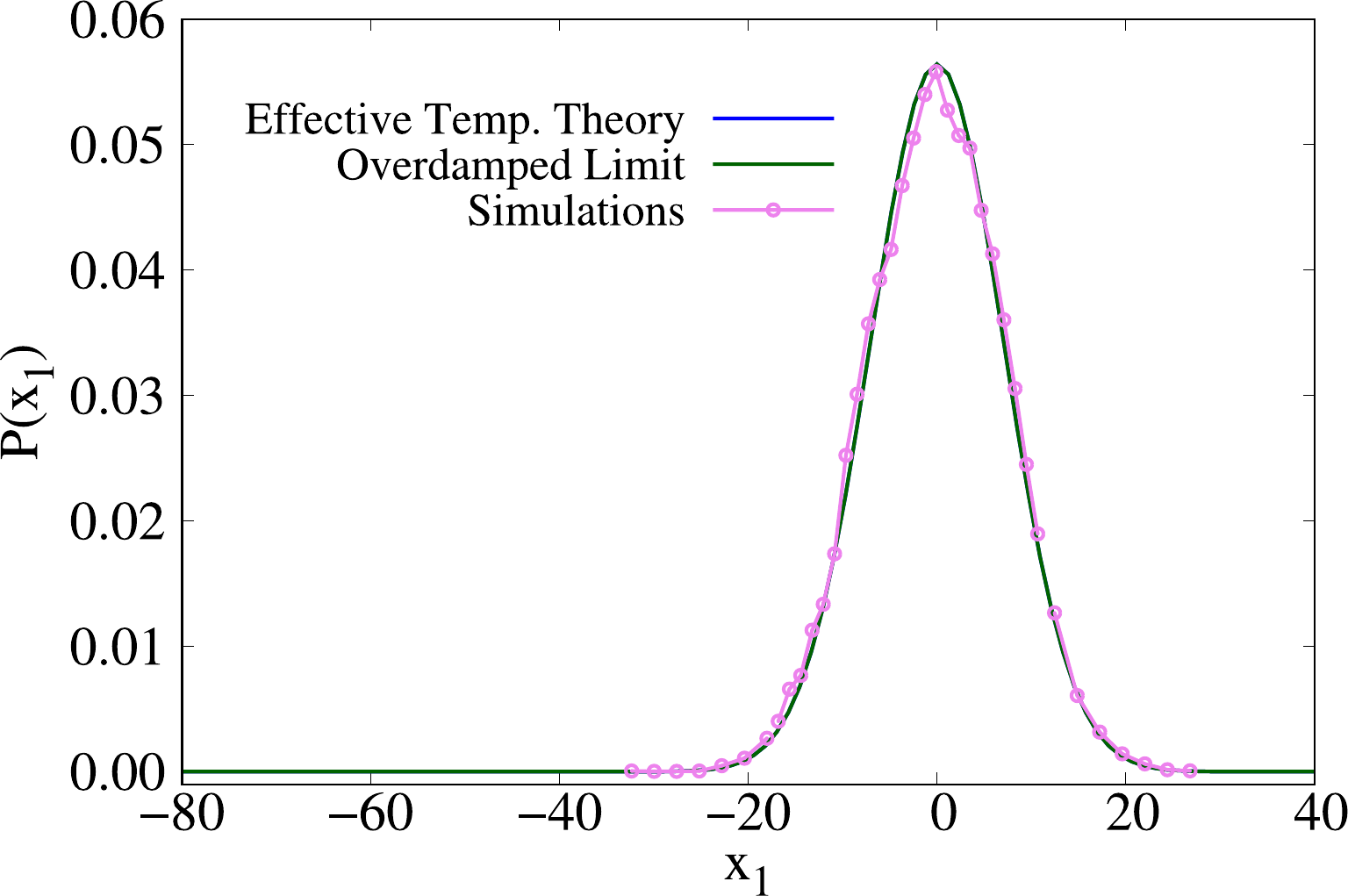}
\caption{The marginal probability distribution ($P(x_{1})$) is derived from the three approaches.  Here, $k=1$, $\gamma_{L} = \gamma_{R}=1$, $T_{L}=99$, and $T_{R}=1$. These results are independent of mass ($m$).}
\label{fig_summary}
\end{figure}
Thus, we observe that the marginal probability distribution for the position ($P(x_{1})$) comes as equal in all the three approaches for the one-bead system, as shown in Fig. (\ref{fig_summary}). 

%
\subsection{\label{Ent_OneBead} The entropy 
production in one-bead system}
\begin{figure}
\includegraphics[scale=0.5]{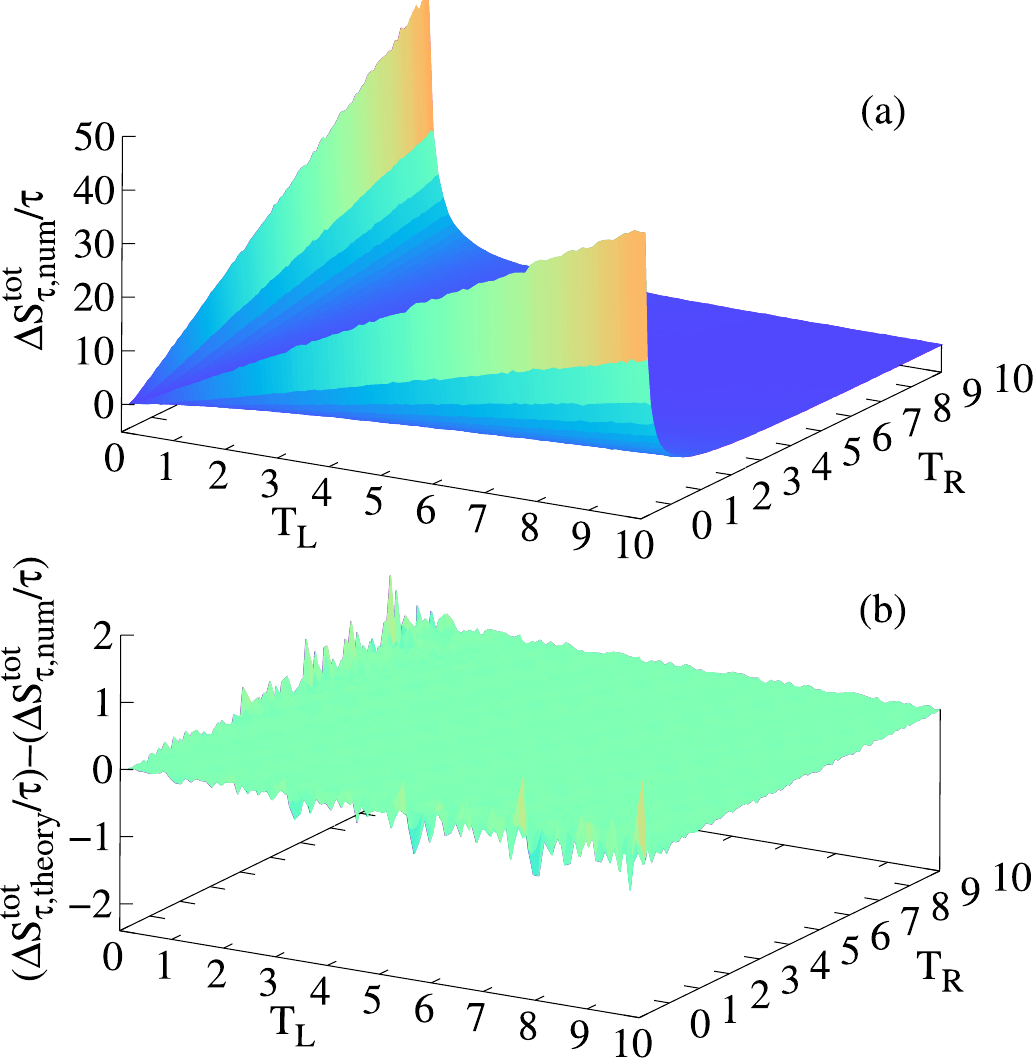}
\caption{(a) The numerically computed rate of entropy production and (b) its deviation from the theoretical prediction [eq. (\ref{EP_Prod_thermo})] versus bath temperatures. The small deviation (one order less) of theoretical prediction from the numerical results  indicates a good match. The simulations assume  
$\gamma_{L}=\gamma_{R}=1$, $k=1$, and $m=1$.}
\label{Therm_Ent_Prod_OneBead}
\end{figure}

In this subsection, we derive the form of entropy production for the one-bead system. From thermodynamics we know, that entropy production is defined as
\begin{equation}
\Delta S_{\tau}^{tot}=-\tau\left ( \frac{\dot{Q}_{L}}{T_{L}} + \frac{\dot{Q}_{R}}{T_{R}} \right)
\label{ent_prod}
\end{equation}
Following \cite{wu2022nonequilibrium}, we can write the heat exchange between a bath and the system as
\begin{gather}
\dot{Q}_{L} = \gamma_{L} \left( \frac{T_{L}}{m}-\langle v_{1}^{2} \rangle \right) \label{heatrate_left},\\
\dot{Q}_{R} = \gamma_{R}\left( \frac{T_{R}}{m}-\langle v_{1}^{2} \rangle \right)\label{heatrate_right}.
\end{gather}
From eqs. (\ref{CovMatrix_OneBead}) and (\ref{cov_matrix}), we get $\langle v_{1}^{2} \rangle = T_{e}/m$ and from eq. (\ref{Teff_OneBead}) $T_{e}=(\gamma_{L}T_{L}+\gamma_{R}T_{R})/(\gamma_{L}+\gamma_{R})$. Substituting this in 
the above, we get
\begin{equation}
  \dot{Q}_{L} = -\dot{Q}_{R} =   
 \frac{\gamma_{L}\gamma_{R}(T_{L}-T_{R})}
 {m(\gamma_{L}+\gamma_{R})}.
 \label{Heat_flux_LR}
\end{equation}
Then  eq. (\ref{ent_prod}) can be expressed as
\begin{gather}
\Delta S_{\tau}^{tot} = 
 \frac{\tau (T_{L}-T_{R})^{2}\gamma_{L}\gamma_{R}}{m(\gamma_{L}+\gamma_{R})T_{L}T_{R}}.
\label{EP_Prod_thermo}
\end{gather}

which looks like Fig. (\ref{Therm_Ent_Prod_OneBead}) for some set of parameters.

\section{\label{SSPD_TwoBead} The two-beads system}
\begin{figure}
\includegraphics[scale=0.4]{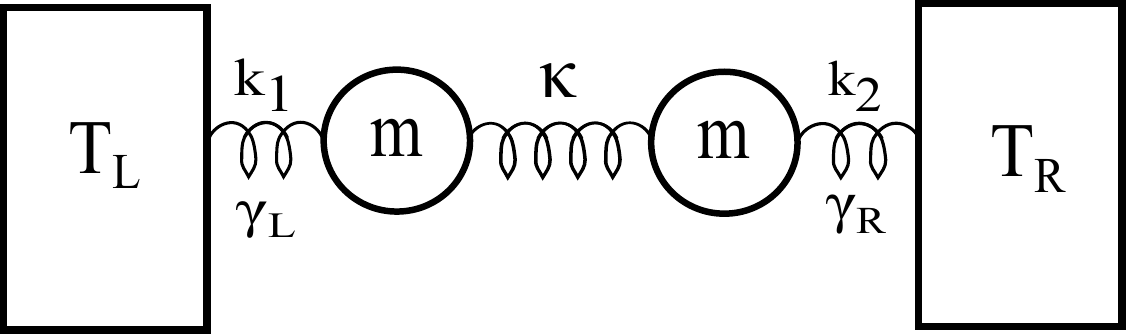}
\caption{Two beads, each of mass $m$, simultaneously 
coupled to each other with a spring having spring constant  $\kappa$ as well as with a bath at temperature $T_{L}$ ($T_{R}$) with spring constant $k_{1}$ ($k_{2}$). The frictional drag coefficient is  $\gamma_{L}$ ($\gamma_{R}$) for the left (right) particle.}
\label{twobeadssetup}
\end{figure}
In a similar manner, we can derive the probability distribution of a two-beads system. As we show below,
the residual matrix comes as non-zero in this case and the effective temperature is a bit complicated. The model system consists of two beads, each directly coupled to a thermal 
reservoir kept at a different temperature (see Fig. (\ref{twobeadssetup})). The coupling to the left bath is 
characterized by the spring constant $k_{1}$ and for the right bath 
by the constanr $k_{2}$. Similarly, the beads are coupled to each other by means of a spring with spring constant $\kappa$. 
 The Langevin equations \cite{saito2007fluctuation, kundu2011large, fogedby2012heat} for this system are 
\begin{align}
\frac{dx_{1}}{dt} = & \;  v_{1}, \label{pos1_Langevin} \\
\frac{dv_{1}}{dt} = & -\frac{k_{1}x_{1}}{m}
+\frac{\kappa(x_{2}-x_{1})}{m} 
-\frac{\gamma_{L}v_{1}}{m}+\xi_{L}(t), \label{vel1_Langevin}\\
\frac{dx_{2}}{dt} = & \;  v_{2}, \label{pos2_Langevin} \\
\frac{dv_{2}}{dt} = & 
-\frac{k_{2}x_{2}}{m}
-\frac{\kappa(x_{2}-x_{1})}{m} 
-\frac{\gamma_{R}v_{2}}{m}+\xi_{R}(t),
 \label{vel2_Langevin}
\end{align}
where $x_{1}$ ($x_{2}$) is the displacement of the left (right) bead from its respective equilibrium position,
and $v_{1}$ ($v_{2}$) its velocity. The white noises due to thermal baths, satisfy $\langle \xi_{\alpha}(t) \rangle = 0$ and $\langle \xi_{\alpha}(t)\xi_{\alpha}(t') \rangle = 2{\gamma_{\alpha}T_{\alpha}}\delta(t-t')/m^2$ with $\alpha=L,R$ and $m$ is the mass of each bead. Here as well, the physical units for the degrees of freedom and system parameters could be appropriately chosen to fit the description of the langevin model (eqs. (\ref{pos1_Langevin}) - (\ref{vel2_Langevin})).

\subsection{\label{CPDA_TwoBead} Comparison of the probability distribution for two approaches, with mass and the overdamped limit}


Here also, the Langevin equations  
(\ref{pos1_Langevin}-\ref{vel2_Langevin}) represent an Ornstein-Uhlenbeck process where  the probability distribution for positions and velocities of left bead ($x_{1}, v_{1}$) and right bead ($x_{2},v_{2}$)  is given with the help of the inverse of the covariance matrix ($\boldsymbol{\sigma}^{-1}$) \cite{van2004microscopic, seifert2012stochastic, barato2015thermodynamic} as follows.
\begin{equation}
P(\boldsymbol{z} = [x_{1},x_{2},v_{1},v_{2}]^{T}) = 
\frac{\exp({-{\boldsymbol{z}^{T}\boldsymbol{\sigma}^{-1}\boldsymbol{z}}/{2}})}{\sqrt{Det[2\pi \boldsymbol{\sigma}]}}
\label{Prob_interms_cov_matrix_2B}
\end{equation}
In Appendix B, we have laid out the formulation for the effective temperature ($T_{e}$) from the Langevin equations for the two-beads system, which is used to write the joint probability distribution in the form of Boltzmann distribution as:  $P(\boldsymbol{z}) = \exp({-(H+\Delta H)/T_{e}})/\mathcal{Z}$, where $H$ and $\Delta H$ are given in eqs. (\ref{Hamiltonian_2B}) and (\ref{dH_2B}) respectively.

\begin{figure}
\includegraphics[scale=0.65]{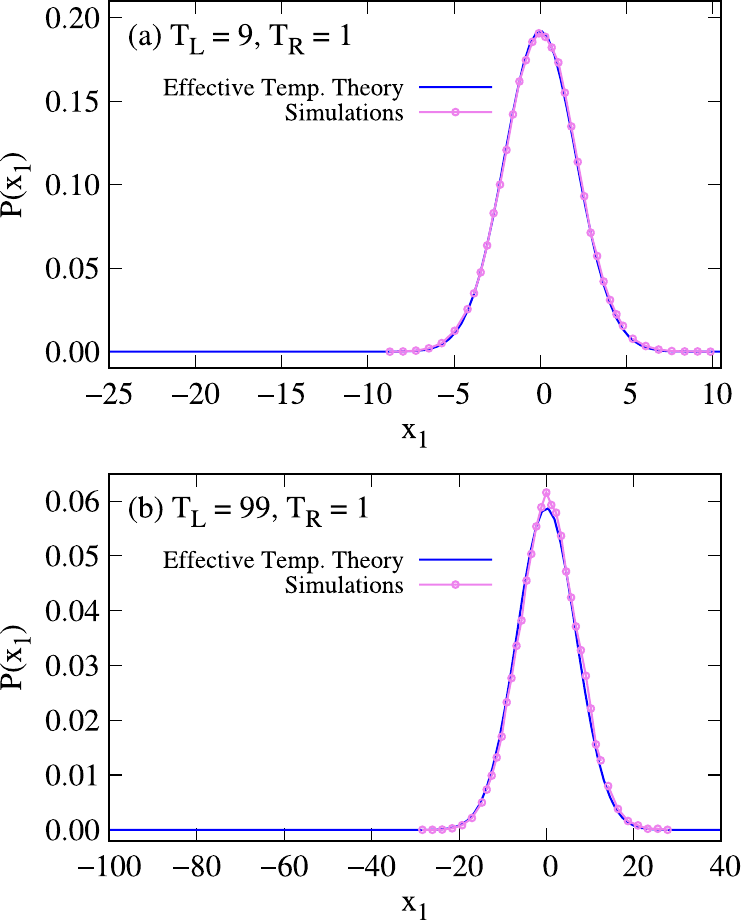}
\caption{The numerically computed marginal probability distribution ($P(x_{1})$) in comparison to the distribution obtained from effective temperature theory.  The simulations assume $T_{L}=9$ (a) and $T_{L}=99$ (b), $T_{R}=1$, $\gamma_{L}=\gamma_{R}=1$, $k_{1}=k_{2}=1$, $\kappa = 2$, and $m=0.01$.}
\label{mar_prob_diffapproach}
\end{figure}

In Fig. (\ref{mar_prob_diffapproach}), we include a comparison plot for the marginal probability distribution ($P(x_{1})$) in the position space of the first bead ($x_{1}$) derived from the $P(\boldsymbol{z})$. The marginal distribution 
is given by $P(x_{1})=\int_{-\infty}^{\infty}dx_{2}\int_{-\infty}^{\infty}dv_{1}\int_{-\infty}^{\infty} dv_{2}P(\boldsymbol{z})$.

\begin{figure}
\includegraphics[scale=0.32, angle=270]{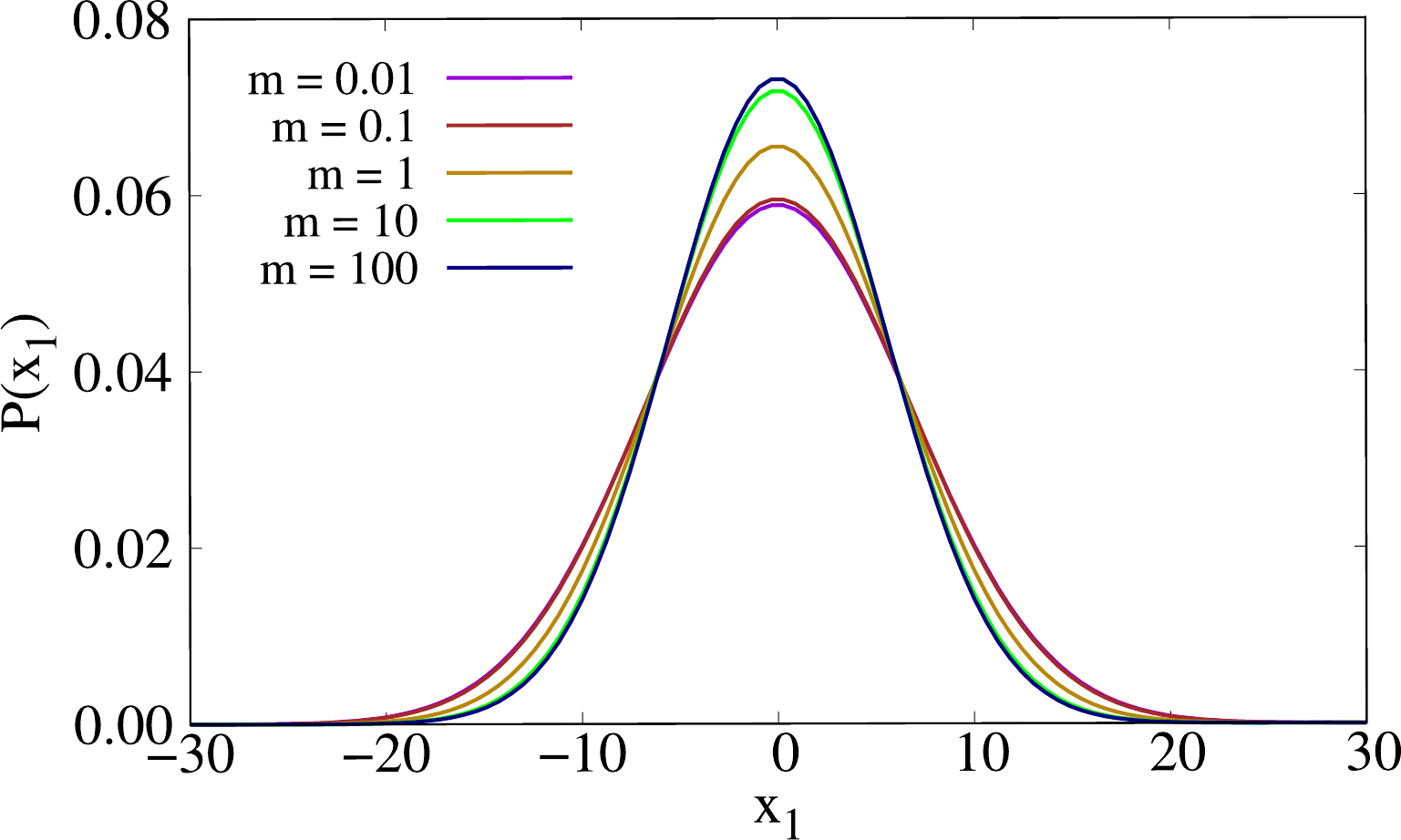}
\caption{The numerically computed marginal probability distribution ($P(x_{1})$) for different values of the mass $m$. The distributions become narrower  with an increase in mass.
Here, $T_{L}=99$, $T_{R}=1$, $\gamma_{L}=\gamma_{R}=1$, $k_{1}=k_{2}=1$, and $\kappa = 2$.}
\label{mar_prob_diffmass}
\end{figure}

\begin{figure}
\includegraphics[scale=0.65]{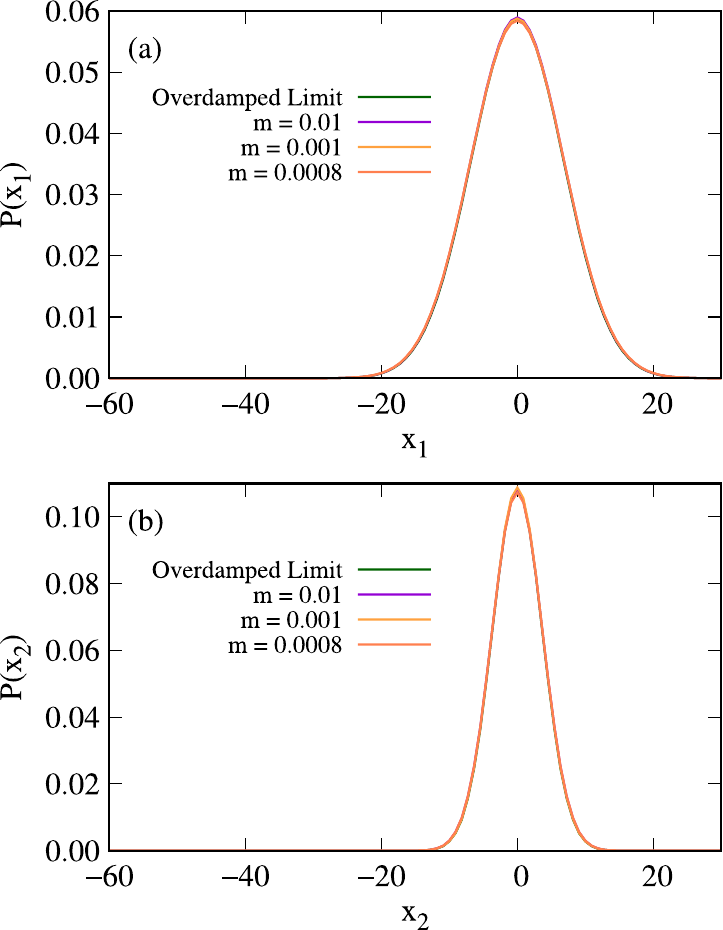}
\caption{The marginal probability distribution for the position ($x_{1}$) of the bead to the left (a) and position ($x_{2}$) of the bead to the right (b). The marginal distributions for the overdamped limit are compared with the numerically obtained distributions of different small masses. The distributions from the overdamped limit and the three small masses lie on top of each other. The simulations were done at  $T_{L}=99$, $T_{R}=1$, $\gamma_{L}=\gamma_{R}=1$, $k_{1}=k_{2}=1$, and $\kappa = 2$.}
\label{Px1x2_od}
\end{figure}

The distributions from the effective temperature overlap with the ones obtained numerically. 
Fig. (\ref{mar_prob_diffmass}) shows the effect of the mass
parameter where we observe the distribution becoming narrower with an increase in mass.
When mass tends to zero, we reach the overdamped limit. The steady-state overdamped probability distribution ($P_{od}(x_{1},x_{2})$) is of the form
\begin{equation}
P_{od}(x_{1},x_{2}) = \frac{1}{2\pi \sqrt{Det[\boldsymbol{\sigma}_{od}]}}\exp^{-\frac{[x_{1},x_{2}]\boldsymbol{\sigma}_{od}^{-1}[x_{1},x_{2}]^{T}}{2}}
\label{Prob_OD_TwoBead}
\end{equation}
where $\boldsymbol{\sigma}_{od}$ satistfies the Lyapunov equation: $\boldsymbol{F.\sigma}_{od}+\boldsymbol{\sigma}_{od}.\boldsymbol{F}^{T} = 2\boldsymbol{D}$ \cite{maes2003time, seifert2008stochastic, parrondo2009entropy, tome2015stochastic}. The form of $\boldsymbol{F}$ could be derived from the Langevin equations (\ref{vel1_Langevin}) and (\ref{vel2_Langevin}) after 
taking the $m \rightarrow 0$ limit. This Langevin equation at the overdamped limit could be compactly written as
\begin{equation}
\frac{d[x_{1},x_{2}]^{T}}{dt} = - \boldsymbol{F}.[x_{1},x_{2}]^{T} + [\xi_{L}(t)/\gamma_{L},\xi_{R}(t)/\gamma_{R}]^{T}. \label{vec_od_Langevin}
\end{equation}
The form of $\boldsymbol{F} =  \begin{bmatrix}
\frac{k_{1}+\kappa}{\gamma_{L}} & -\frac{\kappa}{\gamma_{L}}  \\
-\frac{\kappa}{\gamma_{R}} & \frac{k_{2}+\kappa}{\gamma_{R}}
\end{bmatrix}$ and $\boldsymbol{D} =  \begin{bmatrix}
\frac{T_{L}}{\gamma_{L}} & 0  \\
0 & \frac{T_{R}}{\gamma_{R}}
\end{bmatrix}$. 
By assuming $\boldsymbol{\sigma}_{od} =  \begin{bmatrix}
a & b  \\
b & c
\end{bmatrix}$, 
we solve the Lyapunov equation 
and so obtain
\begin{align}
a &= \frac{T_{L}+b\kappa}{k_{1}+\kappa}, \\
b & =\frac{\kappa \left(\frac{T_{L}}{\gamma_{R}(k_{1}+\kappa)}+\frac{T_{R}}{\gamma_{L}(k_{2}+\kappa)} \right)}{\left( (k_{1}+\kappa)(k_{2}+\kappa)-\kappa^{2} \right)\left(\frac{1}{\gamma_{R}(k_{1}+\kappa)}+\frac{1}{\gamma_{L}(k_{2}+\kappa)} \right)}, \\
c &= \frac{T_{R}+b\kappa}{k_{2}+\kappa}.
\end{align}
We have compared the marginal distributions for positions of both the beads, given by $P(x_{1}) = \int dx_{2}P_{od}(x_{1},x_{2})$ and $P(x_{2}) = \int dx_{1}P_{od}(x_{1},x_{2})$ with the numerically obtained marginal distributions at very small masses as in Fig. (\ref{Px1x2_od}). We see that the distributions are very close  to each other for all the cases, thus satisfying the definition of probability distribution in the  overdamped limit.

\subsection{\label{Ent_TwoBead} The Entropy Production 
in the two-beads system}
\begin{figure}
\includegraphics[scale=0.35, angle=270]{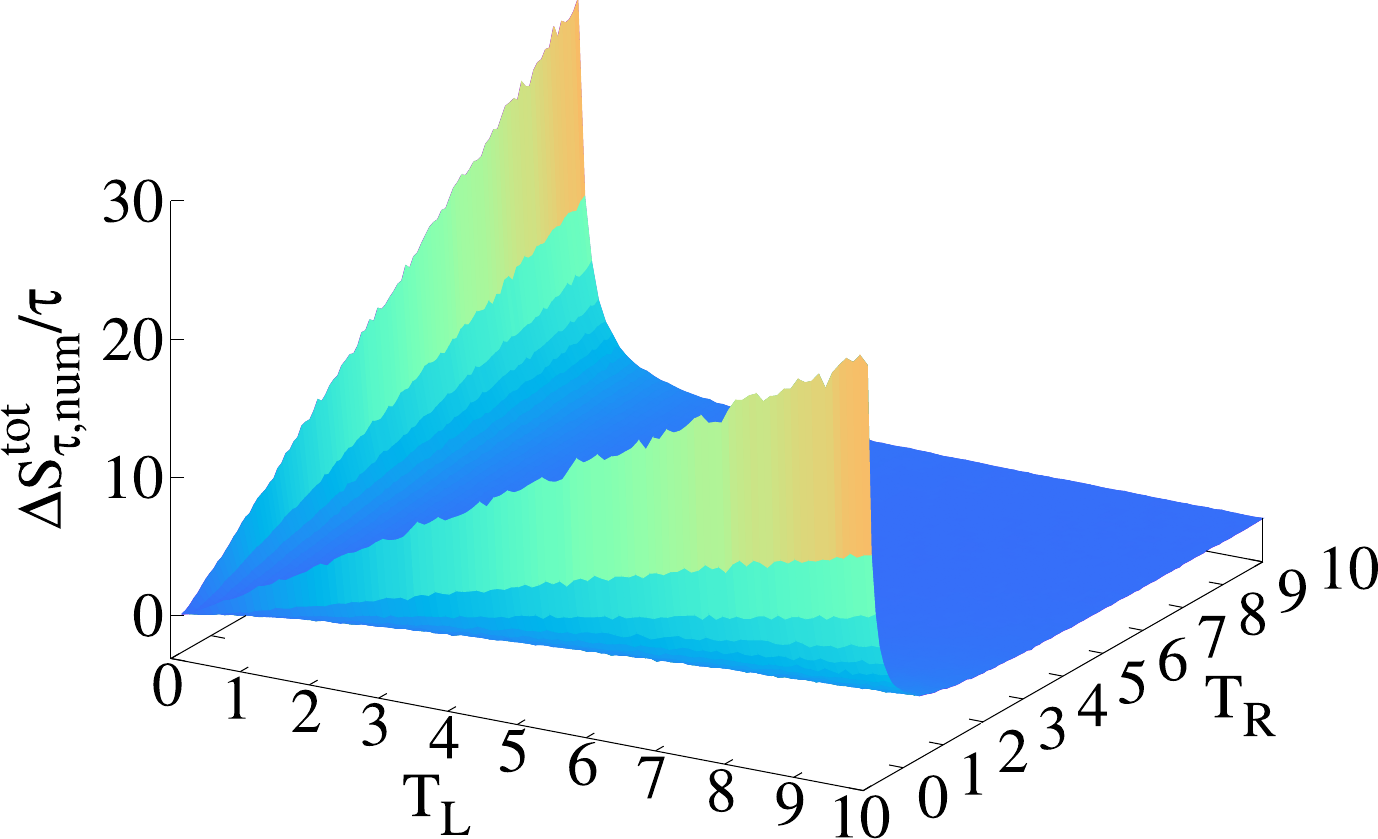}
\caption{The numerically computed rate of entropy production rate for the temperature of the heat bath towards left ($T_{L}$) and the same towards right ($T_{R}$) for two beads and the spring model. We have done the simulations for $\gamma_{L}=\gamma_{R}=1$, $k_{1}=k_{2}=1$, $\kappa = 2$, and $m=1$.}
\label{Therm_Ent_Prod}
\end{figure}

We can define the entropy produced as in 
eq. (\ref{ent_prod}) and shown in Fig. (\ref{Therm_Ent_Prod}),
where 
\begin{gather}
\dot{Q}_{L} = \gamma_{L} \left( {T_{L}}/{m}-\langle v_{1}^{2} \rangle \right) \label{heatrate_left_2Bead}\\
\dot{Q}_{R} = \gamma_{R}\left( {T_{R}}/{m}-\langle v_{2}^{2} \rangle \right)\label{heatrate_right_2Bead}
\end{gather}
Thus, the entropy production here behaves similarly to that 
in one-bead system. Even though the rate of entropy production is reduced to almost half in comparison to the one-bead system,  we still have a symmetry about $T_{L}=T_{R}$ line (see Fig. (\ref{Therm_Ent_Prod})) and the entropy production is zero along $T_{L}=T_{R}$ line. We also see the rate of entropy production increasing with increase in difference of temperature of the two heat reservoirs.

\section{\label{ergodicity} Ergodicity}
\begin{figure}
\includegraphics[scale=0.3]{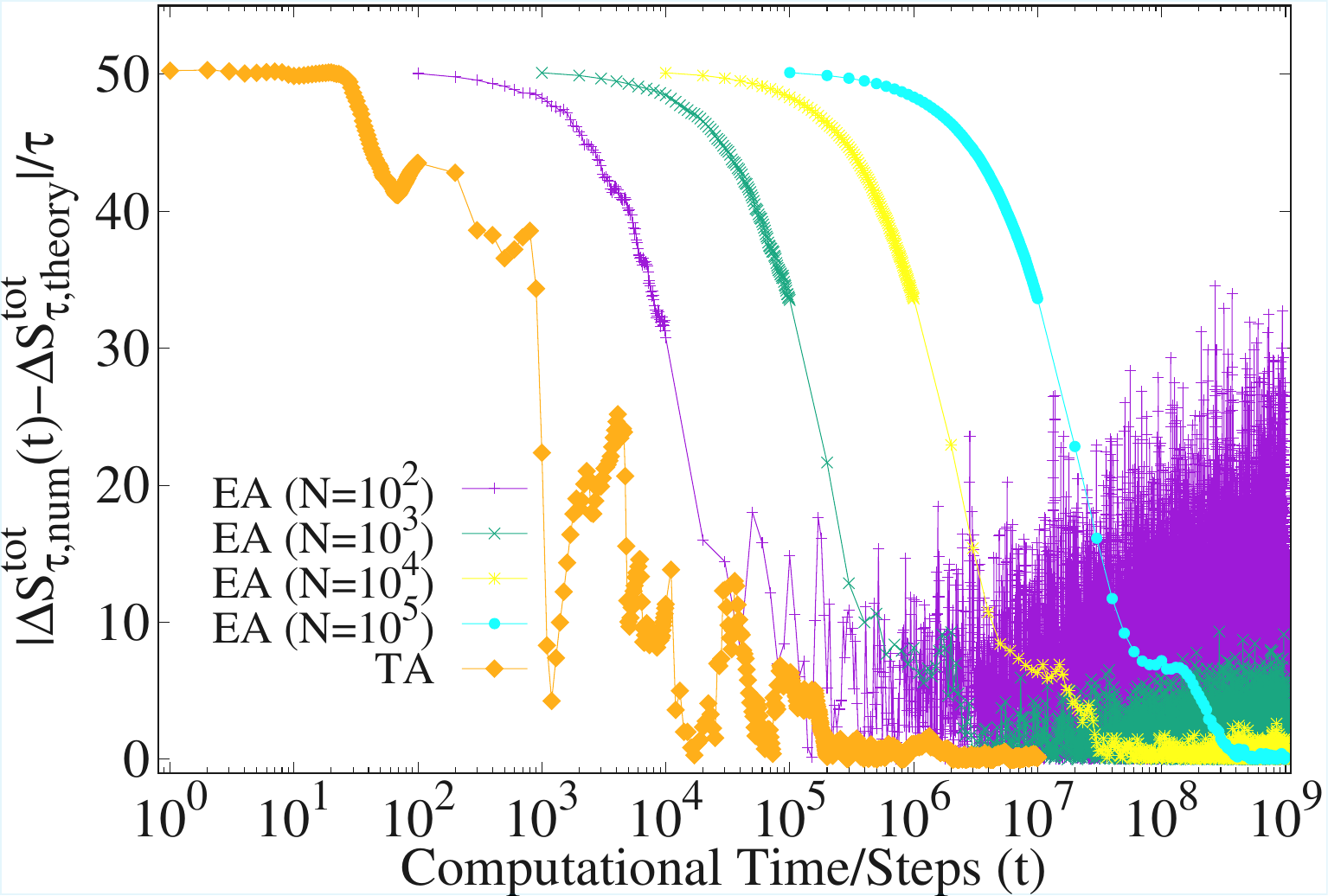}
\caption{The absolute value of deviation of rate of entropy production obtained at each computational time or step from the theoretical value for a one bead system. The curves are computed from ensemble average (EA) with $N$ number of samples or trajectories and time average (TA). We see the curve for time average (orange) reaching steady states fastest and hence small deviation from the theoretical value after approximately $10^{6}$ time steps. The curves for ensemble average reach steady states at a later stage, increasing time with increasing the number of trajectories. But, the fluctuations become significantly smaller for the largest $N=10^{5}$ curve (cyan) which reach steady state at about $5 \times 10^{8}$ time steps. We have done the simulations for $T_{L}=99$, $T_{R}=1$, $\gamma_{L}=\gamma_{R}=1$, $k_{1}=k_{2}=1$, $\kappa = 2$, and $m=1$.}
\label{EnsvsTS_Ent_Prod}
\end{figure}

The one-bead and the two-beads systems are simple systems for studying non-equilibrium processes. But, these systems behave like equilibrated systems which could be achieved by keeping the temperature of both the heat baths equal. It would be interesting to study these systems' ergodic properties. It would be a valuable property to study as the computational time taken for reaching the steady states in the ensemble averaging of quantities computed with $N$ configurations of the system is more for reaching the steady states compared to the time averaging of the same quantity. In our case, for ensemble averaging, we took $10^{5}$ configurations of the one-bead or the two-beads set up. We evolved them using the Langevin equations using Euler-Maruyama method \cite{nayak2021numerical, erfanian2016using, bayram2018numerical} from initial conditions ($(x_{1},v_{1})=(0,0)$ and $(x_{2},v_{2})=(0,0)$) till it reaches steady state. The computational time steps ($t$) used in this section, are given in units of time taken for completing one langevin step in incrementing position and velocity due to noise in the computer. The $dt$ used in integrating the Langevin equations for one bead setup (eqs. \ref{pos_Langevin_1Bead} and \ref{vel_Langevin_1Bead}) and two bead setup (eqs. \ref{pos1_Langevin} to \ref{vel2_Langevin}) is $10^{-3}$ units. We did the ensemble averaging after $10^{4}$ steps of this collection of $N=10^{5}$ configurations. The cyan curve in Fig. (\ref{EnsvsTS_Ent_Prod}) represents this case for one-bead system reaching the steady state at the longest time ($t \approx 5 \times 10^{8}$), but with the least fluctuations as compared to the other curves obtained from ensemble averaging for a smaller number of trajectories or configurations even though they appear to have reached steady state earlier.  In order to calculate the time averages, we used a single configuration having all positions and velocities to be zero and made it evolve using the Langevin equations corresponding to the one-bead or the two-beads system. We used upto $10^{7}$ configurations for the averaging which is just summing the quantities for each configuration and dividing by the number of configurations used in the sampling. The computational time taken for averaging through the ensemble approach took two orders more than the averaging done through the time series for our case. Because, as seen by the orange curve, the time average in Fig. (\ref{EnsvsTS_Ent_Prod}) reaches steady state with least fluctuations much before ($t \approx 10^{6}$) than the other curves from ensemble averaging. We took samples till $t=10^{7}$ time steps for taking the time average of quantities which is two orders lower than the computational time for the cyan curve ($t = N\times10^{4} = 10^{9}$). Therefore, the probability distributions and the rate of entropy production studied throughout this paper are derived from time series having $10^{7}$ samples and the average is taken over this time series as this would need a smaller computational time.

\begin{figure}
\includegraphics[scale=0.5]{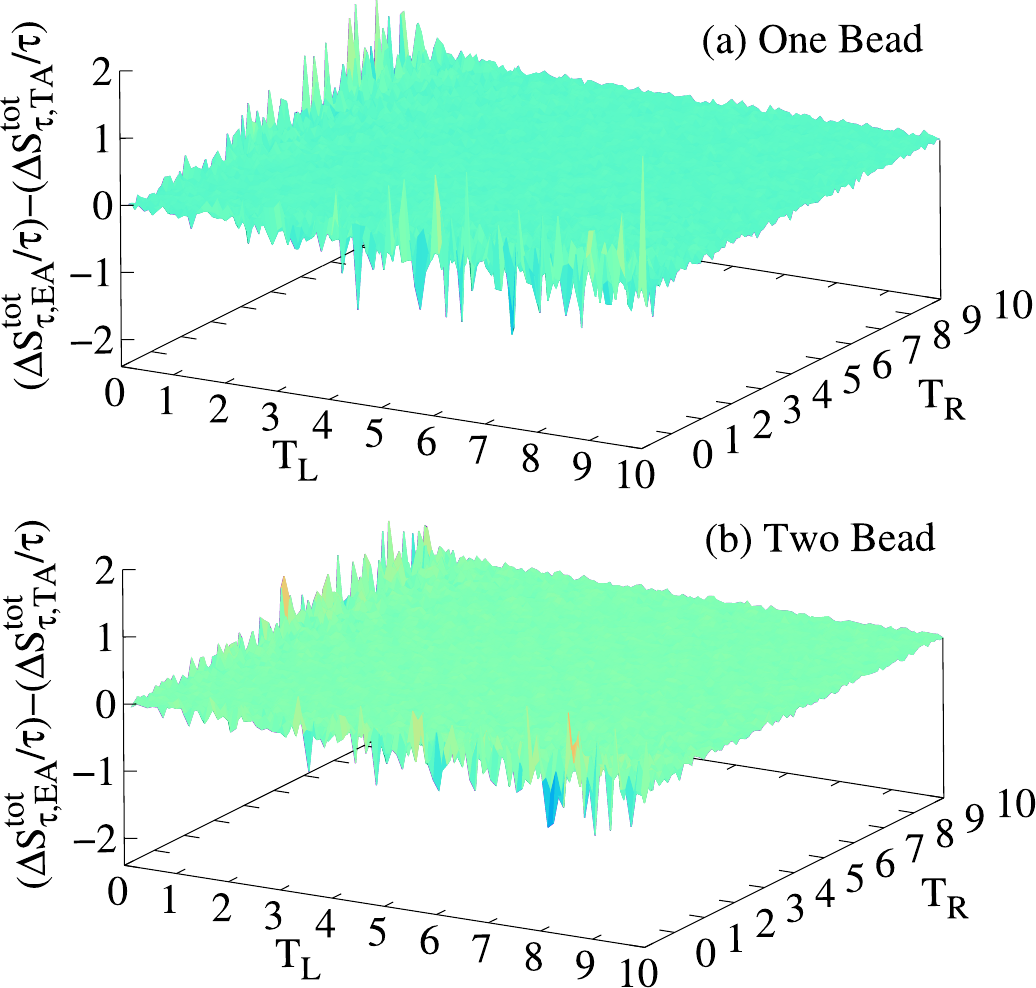}
\caption{The difference of rate of entropy production for (a) one bead system obtained by taking the average of $\langle v_{1}^{2} \rangle$ in eqn's. (\ref{heatrate_left}) and (\ref{heatrate_right}) using ensemble and time series. The simulations for one bead setup are done for $T_{L}=99$, $T_{R}=1$, $\gamma_{L}=\gamma_{R}=1$, $k=1$, and $m=1$. The same difference for (b) two bead system obtained by averaging $\langle v_{1}^{2} \rangle$ and $\langle v_{2}^{2} \rangle$ in eqn's (\ref{heatrate_left_2Bead}) and (\ref{heatrate_right_2Bead}) respectively using ensemble and time series. The difference in both cases is negligible implying ergodicity. The simulations for two bead setup are done for $T_{L}=99$, $T_{R}=1$, $\gamma_{L}=\gamma_{R}=1$, $k_{1}=k_{2}=1$, $\kappa = 2$, and $m=1$.}
\label{Dev_Ent_Prod_EAandTA}
\end{figure}

We have computed the rate of entropy production ($\Delta S^{tot}_{\tau}/\tau$) for the two set ups, using the ensemble 
average (EA) and the time average  (TA) and thus  illustrate the property of ergodicity in both the set ups in Fig. (\ref{Dev_Ent_Prod_EAandTA}). The difference between the rate of entropy production obtained from ensemble averaging and the time averaging is negligible which is one order less as compared to the values of rate of entropy production. 

\begin{figure}
\includegraphics[scale=0.35, angle=270]{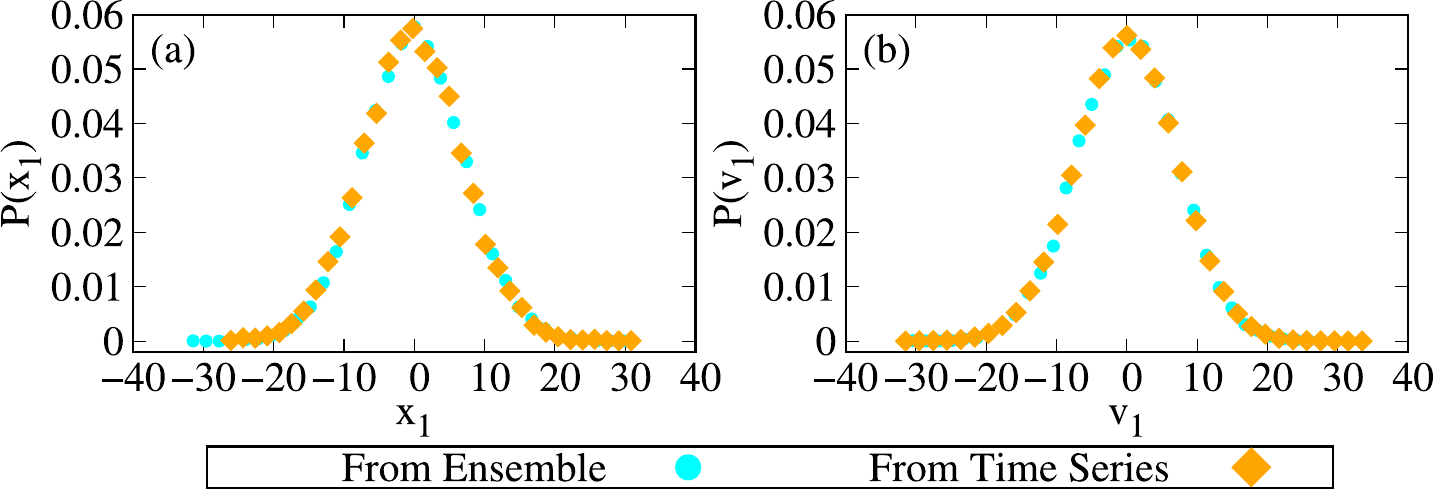}
\caption{The probability distribution for position (a) and velocity (b) of the bead in one bead setup. The difference between the two curves in a plot is negligible implying that the distribution obtained from elements in an ensemble and the time series results in the same distribution proving ergodicity. The simulations are done for $T_{L}=99$, $T_{R}=1$, $\gamma_{L}=\gamma_{R}=1$, $k=1$, and $m=1$.}
\label{Prob_OneBead_EAandTA}
\end{figure}

\begin{figure}
\centering{\includegraphics[scale=0.35]{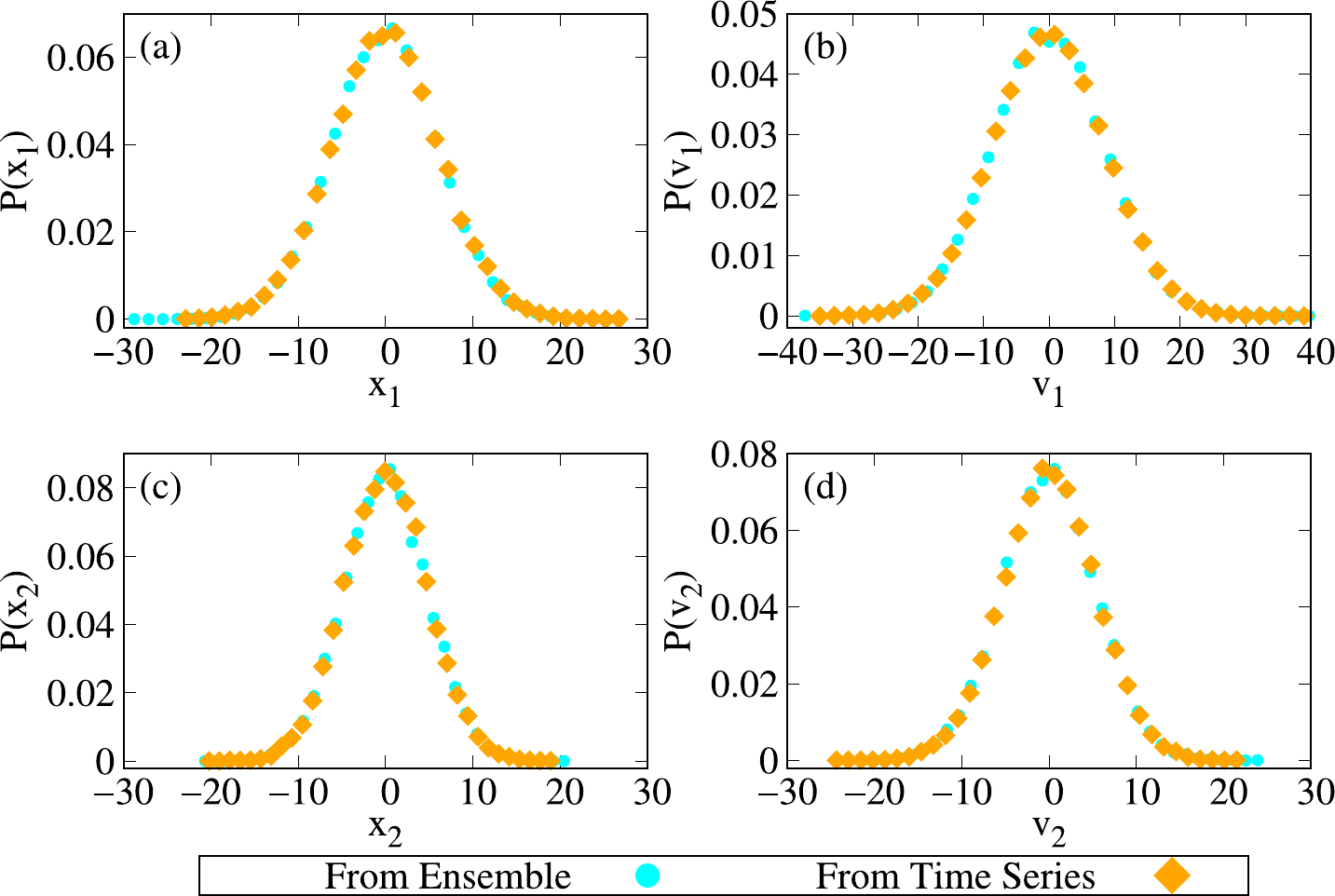}}
\caption{The probability distribution for the position (a), velocity (b) of the leftmost bead and the position (c), velocity (d) of the rightmost bead in two bead setups. The difference between the two curves in a plot is negligible implying that the distribution obtained from elements in an ensemble and the time series results in the same distribution proving ergodicity.The simulations are done for $T_{L}=99$, $T_{R}=1$, $\gamma_{L}=\gamma_{R}=1$, $k_{1}=k_{2}=1$, $\kappa = 2$, and $m=1$.}
\label{Prob_TwoBead_EAandTA}
\end{figure}

We obtain a list of positions ($x_{1}$) and velocities ($v_{1}$) for the one-bead set up from the configurations in an ensemble or the time series and from these we can make a histogram or marginal probability distributions for the different degrees of freedom. These marginal distributions are shown in Fig. (\ref{Prob_OneBead_EAandTA}). We observe that the distributions obtained from both procedures lie on top of each other. This means that the probability distribution during the time series and those contained in the ensemble are equal resulting in a much more stronger proof of ergodicity.
We also obtain the list of positions ($x_{1}$,$x_{2}$) and velocities ($v_{1}$,$v_{2}$) for the two-beads set up from the ensemble and the time series and have made the marginal distributions for all the degrees of freedom. These marginal distributions are shown in Fig. (\ref{Prob_TwoBead_EAandTA}). Similar to one-bead set up, we see that the distributions obtained from both the ensemble and time series for the two beads lie on top of each other. 
The evidences given here in this section about the time average being equal to the ensemble average for the rate of entropy production in Fig. (\ref{Dev_Ent_Prod_EAandTA}), the marginal distributions obtained for all the degrees of freedom matching for data from time series and many realizations at fixed time and in gaussian processes like these where the probability distribution converging to a stationary distribution having unique covariance matrix at steady states for all practical purposes shows that both our systems satisfy the criterion of ergodicity.
For our specific systems, ergodicity can be shown, because their dynamics are governed by a linear Langevin equation with additive Gaussian noise and the damping or the noise satisfies fluctuation-dissipation. Thus, ergodicity is expected and often provable, especially in the underdamped regime \cite{frank2005nonlinear, eckmann1999non, risken1996Fokker, maes2003time}.

\section{\label{Conclusions} Conclusions}
In this paper, we were able to introduce an effective temperature which enables us to write the joint probability distributions for position and velocity as a Boltzmann weight with this effective temperature acting as the temperature and effective hamiltonian. Thus, we were able to cast a non-equilibrium problem to an equilibrium one for underdamped systems. We have shown this approach applies to one and two-bead spring systems. This probability distribution was also obtained from numerical simulations. These numerical simulations involved solving the Langevin equation given in eq. (\ref{pos_Langevin_1Bead}) and eq. (\ref{vel_Langevin_1Bead}) for one bead and spring model and eq. (\ref{pos1_Langevin}) to eq. (\ref{vel2_Langevin}) for two bead and spring models. We were able to reproduce the probability distributions from both the approaches. And, we derived the marginal probability distribution in position space from them. For one bead problem, we got a distribution independent of mass, and hence these marginal distributions were equal to the distribution obtained from the overdamped limit. For the two bead problem also, the marginal distributions could be obtained from the joint probability distribution which is exactly obtained using the effective temperature as shown in Appendix B. Unlike the one-bead problem, the marginal distributions depend on mass for the two-bead case. We compared these distributions to that obtained from the overdamped limit and see them becoming equal to the latter as the mass becomes really small. We see this exactness for both the marginal distributions in the position of the leftmost bead as well as the rightmost bead. We also calculate the rate of entropy production for the one-bead and two-bead systems and have described their features and could possibly in future, study the evolution of various quantities similar to rate of entropy production as the bead spring models approach the non-equilibrium steady states in lines similar to work of Netz \cite{netz2020approach}. We have studied and got the time taken to reach steady state for a protocol using time averaging is two orders less than ensemble averaging and equivalence of rate of enropy production and probability distributions suggesting ergodicity to be followed by these systems as well.

\section{Appendix A}
The Langevin equations (\ref{pos_Langevin_1Bead}) and  (\ref{vel_Langevin_1Bead}) can be written compactly as
\begin{equation}
\frac{d\boldsymbol{z}}{dt} = - \boldsymbol{Az} + \boldsymbol{\xi}. \label{vec_Langevin_1Bead}
\end{equation}
For one-bead system, $\boldsymbol{z} = [x_{1},v_{1}]^{T}$ and $\boldsymbol{\xi} = [0, \xi_{L} +\xi_{R}]^{T}$ where $T$ represents the transpose operation. The coefficient matrix 
is written as
\begin{equation}
\boldsymbol{A} =
\begin{bmatrix}
0 & -1 \\
k/m & (\gamma_{L}+\gamma_{R})/m
\end{bmatrix}.
\label{coeff_matrix_1Bead}
\end{equation}
 The covariance matrix ($\boldsymbol{\sigma}$)
 is given as
\begin{equation}
\boldsymbol{\sigma} =
\langle \boldsymbol{zz}^{T} \rangle_{t \rightarrow \infty} =   \begin{bmatrix}
\sigma_{x_{1}x_{1}} & \sigma_{x_{1}v_{1}} \\
\sigma_{v_{1}x_{1}} & \sigma_{v_{1}v_{1}}
\end{bmatrix},
\label{cov_matrix}
\end{equation}
where 
$\sigma_{x_{1}x_{1}} = \langle x_{1}x_{1} \rangle_{t \rightarrow \infty}$, $\sigma_{v_{1}v_{1}}=\langle v_{1}v_{1} \rangle_{t \rightarrow \infty}$ and 
$\sigma_{x_{1}v_{1}} = \sigma_{v_{1}x_{1}} = \langle x_{1}v_{1} \rangle_{t \rightarrow \infty}$.

For an Ornstein-Uhlenbeck process described by eq. (\ref{vec_Langevin_1Bead}), the covariance matrix satisfies the continuous time Lyapunov equation \cite{herzel1991risken, gardiner1985handbook}, which means
\begin{equation}
\boldsymbol{A\sigma}+\boldsymbol{\sigma A}^{T} = 2\boldsymbol{D}.
\label{CT_Lyapunov}
\end{equation}
For the one-bead system, the diffusion matrix $\boldsymbol{D}$ 
is written as 

\begin{equation}
\boldsymbol{D} =
\begin{bmatrix}
0 & 0 \\
0 & ({\gamma_{L}T_{L}+
\gamma_{R}T_{R}})/{m^2}
\end{bmatrix}.
\label{Dt_matrix}
\end{equation}
%
For $T_L = T_R = T$,  
the system reaches equilibrium at the temperature $T$, implying that $k\sigma_{x_{1}x_{1}}=m\sigma_{v_{1}v_{1}}=T$ 
while $\sigma_{x_{1}v_{1}}=\sigma_{v_{1}x_{1}}=0$
\cite{van1992stochastic, sekimoto2010stochastic, seifert2012stochastic, tome2015stochastic}.
Next, we define a modified covariance matrix ($\boldsymbol{\Tilde{\sigma}}$) of the form
\begin{equation}
\Tilde{\boldsymbol{{\sigma}}} =
\begin{bmatrix}
\sigma_{x_{1}x_{1}}k & \sigma_{x_{1}v_{1}}m \\
\sigma_{v_{1}x_{1}}k & \sigma_{v_{1}v_{1}}m
\end{bmatrix},
\label{SigmaTilde_matrix}
\end{equation}
such that the equilibrium covariance matrix $\Tilde{\boldsymbol{{\sigma}}}_{\rm eq}=T\boldsymbol{I}_{2}$, with $\boldsymbol{I}_{2}$ being a $2 \times 2$ identity matrix. This motivates us to come up with the following ansatz on non-equilibrium steady states: the revised covariance matrix may be decomposed into a diagonal matrix plus a residual matrix. Tu \cite{tu2025weighted} introduced an effective temperature
$T_e$ for these states, so that
the covariance matrix looks like
\begin{equation}
\Tilde{\boldsymbol{\sigma}} = T_{e}\boldsymbol{I}_{2} + \Tilde{\boldsymbol{\sigma}}^{r},
\label{Modified_Covariance_Matrix_1D}
\end{equation}
where the residual matrix is traceless, i.e. $\text{Tr}\tilde{\boldsymbol{\sigma}}^{r}=0$. 
We can write $\tilde{\boldsymbol{\sigma}} = \boldsymbol{\sigma} \tilde{\boldsymbol{K}}$, where $\boldsymbol{\tilde{K}}$ 
is given by
\begin{equation}
\Tilde{\boldsymbol{K}} =
\begin{bmatrix}
k & 0 \\
0 & m
\end{bmatrix}
\label{KTilde_matrix}.
\end{equation}
Similarly, we define $\tilde{\boldsymbol{A}}$
as 
\begin{equation}
\tilde{\boldsymbol{A}} = \tilde{\boldsymbol{K}}\boldsymbol{A} =
\begin{bmatrix}
0 & -k \\
k & \gamma_{L}+\gamma_{R}
\end{bmatrix}.
\label{ATilde_matrix}
\end{equation}
From eqs. (\ref{CT_Lyapunov}) and  (\ref{Modified_Covariance_Matrix_1D}),
we can write 
\begin{gather}
\tilde{\boldsymbol{A}}\tilde{\boldsymbol{\sigma}}^{r}+\tilde{\boldsymbol{\sigma}}^{r T} \tilde{A}^{T} = 
2 \begin{bmatrix}
0 & 0 \\
0 &  c
\end{bmatrix},
\label{CT_Lyapunov_w_Atilde_1Bead}
\end{gather}
where $c= (\gamma_{L}T_{L}+\gamma_{R}T_{R})-T_{e}(\gamma_{L}+\gamma_{R})$.
Eq. (\ref{CT_Lyapunov_w_Atilde_1Bead}) is a linear equation and could be expressed as a linear combination of a single base $\tilde{\boldsymbol{\sigma}}_{\alpha}$ which satisfies
\begin{equation}
\tilde{\boldsymbol{A}}\tilde{\boldsymbol{\sigma}}_{\alpha}+\tilde{\boldsymbol{\sigma}}_{\alpha}^{T} \tilde{\boldsymbol{A}}^{T} = 2\begin{bmatrix}
0 & 0 \\
0 &  1
\end{bmatrix}.
\label{CT_Lyapunov_w_base}
\end{equation}
Thus, $\tilde{\boldsymbol{\sigma}}^{r} = c \tilde{\boldsymbol{\sigma}}_{\alpha}$. 
Further, $\text{Tr}\tilde{\boldsymbol{\sigma}}^{r}=0$ implies $c=0$ as $\text{Tr}\tilde{\boldsymbol{\sigma}}_{\alpha} \neq 0$. From this constraint, we get 
\begin{equation}
    T_{e} = \frac{\gamma_{L}T_{L}+\gamma_{R}T_{R}}{\gamma_{L}+\gamma_{R}},
\label{Teff_OneBead}    
\end{equation}
     and $\tilde{\boldsymbol{\sigma}}^{r}=c\tilde{\boldsymbol{\sigma}}_{\alpha}=\boldsymbol{0}$, where $\boldsymbol{0}$ is $2\times2$ null matrix.

The steady-state distribution for the one-bead setup is written as 
\begin{equation}
P(\boldsymbol{z}) = 
\frac{1}{\mathcal{Z}}e^{-(H+\Delta H)/T_{e}},
\label{Pz}
\end{equation}
where the Hamiltonian of this system is
\begin{equation}
H = \frac{1}{2}mv_{1}^{2} + \frac{1}{2}k x_{1}^{2} = \frac{1}{2}\boldsymbol{z}^{T}\tilde{\boldsymbol{K}}\boldsymbol{z},
\label{Hamiltonian_1B}
\end{equation}
and 
\begin{equation}
\Delta H = \frac{1}{2}\boldsymbol{z}^{T}\tilde{\boldsymbol{K}}[(\boldsymbol{I}_{2}+\tilde{\boldsymbol{\sigma}}^{r}/T_{e})^{-1}-\boldsymbol{I}_{2}]\boldsymbol{z}.
\label{dH_1B}
\end{equation}
Due to $\tilde{\boldsymbol{\sigma}}^{r}=\boldsymbol{0}$, 
we have $\Delta H = 0$. 
The partition function ($\mathcal{Z}$) appearing  
in $P(\boldsymbol{z})$ above is evaluated
as $2\pi T_{e}/\sqrt{km}$.

\section{Appendix B}
The Langevin equations 
(\ref{pos1_Langevin})-(\ref{vel2_Langevin}) could be
compactly  written as
\begin{equation}
\frac{d\boldsymbol{z}}{dt} = - \boldsymbol{Az} + \boldsymbol{\xi}, \label{vec_Langevin}
\end{equation}
where for the two-bead system  $\boldsymbol{z} = [x_{1},x_{2},v_{1},v_{2}]^{T}$ and $\boldsymbol{\xi} = [0, 0, \xi_{L}, \xi_{R}]^{T}$. The coefficient matrix is 
given by
\begin{equation}
\boldsymbol{A} =
\begin{bmatrix}
0 & 0 & -1 & 0 \\
0 & 0 & 0 & -1 \\
k_{1}/m & -\kappa/m & \gamma_{L}/m & 0\\
-\kappa/m & k_{2}/m & 0 & \gamma_{R}/m\\
\end{bmatrix}.
\label{coeff_matrix}
\end{equation}
 The effective spring constants $k_{1}+\kappa$ will be denoted as $k_{1}$ for the leftmost particle and $k_{2}+\kappa$ as $k_{2}$ for the rightmost particle for the discussion in Appendix B.

Just like the one bead system, here also, the covariance matrix ($\boldsymbol{\sigma}$) will be a symmetric $4 \times 4$ matrix. And the modified matrices $\tilde{\boldsymbol{\sigma}}$ 
and $\tilde{\boldsymbol{A}}$ will satisfy the similar properties as in the one-bead system. Thus,
\begin{equation}
\tilde{\boldsymbol{\sigma}} = \boldsymbol{\sigma} \tilde{\boldsymbol{K}} = \boldsymbol{\sigma}\begin{bmatrix}
\boldsymbol{K} & \boldsymbol{0}\\
\boldsymbol{0} & m\boldsymbol{I}_{2}
\end{bmatrix} =  \begin{bmatrix}
\boldsymbol{\sigma_{xx}}\boldsymbol{K} & \boldsymbol{\sigma_{xv}}m\\
\boldsymbol{\sigma_{vx}}\boldsymbol{K} & \boldsymbol{\sigma_{vv}}m
\end{bmatrix},
\label{Sigmatilde}
\end{equation}
where $\boldsymbol{v} = [v_{1},v_{2}]^{T}$ and $\boldsymbol{x} = [x_{1},x_{2}]^{T}$. 
Similarly, 
\begin{equation}
\tilde{\boldsymbol{A}} = 
\tilde{\boldsymbol{K}}\boldsymbol{A} = 
\begin{bmatrix}
\boldsymbol{0} & -\boldsymbol{K}\\
\boldsymbol{K} & \boldsymbol{\Gamma}
\end{bmatrix} 
\label{Atilde}
\end{equation}
where $\boldsymbol{K} =  \begin{bmatrix}
k_{1} & -\kappa\\
-\kappa & k_{2}
\end{bmatrix}$ and $\boldsymbol{\Gamma} =  \begin{bmatrix}
\gamma_{L} & 0\\
0 & \gamma_{R}
\end{bmatrix}$.
$\tilde{\boldsymbol{\sigma}}$ is assumed to satisfy the following relation \cite{tu2025weighted}
\begin{equation}
\tilde{\boldsymbol{\sigma}} = T_{e}\boldsymbol{I}_{4} + \tilde{\boldsymbol{\sigma}}^{r},
\label{sigmatilde}
\end{equation}
where $\text{Tr}\tilde{\boldsymbol{\sigma}}^{r}=0$ and $\boldsymbol{I}_{4}$ is the $4\times4$ identity matrix. 
Again, from the Lyapunov equation for the two-beads setup 
and the ansatz [eq. (\ref{sigmatilde})], 
we find that the residual matrix must satisfy
\begin{gather}
\tilde{\boldsymbol{A}}\tilde{\boldsymbol{\sigma}}^{r}+\tilde{\boldsymbol{\sigma}}^{r T} \tilde{A}^{T} =  2(\tilde{\boldsymbol{K}}.\boldsymbol{D}.\tilde{\boldsymbol{K}}-T_{e}\boldsymbol{\Gamma}) \nonumber \\
 = 2\begin{bmatrix}
0 & 0 & 0 & 0\\
0 & 0 & 0 & 0\\
0 & 0 & \gamma_{L}(T_{L}-T_{e}) & 0 \\
0 & 0 & 0 & \gamma_{R}(T_{R}-T_{e})
\end{bmatrix},
\label{CT_Lyapunov_w_Atilde}
\end{gather}
where $\boldsymbol{D}$ is the diffusion matrix for 
the two-beads system. The linear equation (\ref{CT_Lyapunov_w_Atilde}) could be expressed as a linear combination of the two bases $\tilde{\boldsymbol{\sigma}}_{\alpha}$ where $\alpha = L, R$ which satisfy 
\begin{equation}
\tilde{\boldsymbol{A}}\tilde{\boldsymbol{\sigma}}_{\alpha}+\tilde{\boldsymbol{\sigma}}_{\alpha}^{T} \tilde{\boldsymbol{A}}^{T} = 2\begin{bmatrix}
0 & 0 & 0 & 0\\
0 &  0 & 0 & 0\\
0 & 0 & \delta_{\alpha L} & 0 \\
0 & 0 & 0 & \delta_{\alpha R}
\end{bmatrix}.
\label{CT_Lyapunov_w_base}
\end{equation}
To obtain $\tilde{\boldsymbol{\sigma}}_{\alpha}$
from the above equation, we express them
in the following form:
\begin{equation}
\tilde{\boldsymbol{\sigma}}_{\alpha} = \begin{bmatrix}
\boldsymbol{B}_{\alpha} & \boldsymbol{J}_{\alpha} \\
\boldsymbol{F}_{\alpha} &  \boldsymbol{G}_{\alpha}
\end{bmatrix},
\label{matrixform_base}
\end{equation}
where  $\boldsymbol{B}_{\alpha}$, $\boldsymbol{F}_{\alpha}$, $\boldsymbol{J}_{\alpha}$ and $\boldsymbol{G}_{\alpha}$ are $2\times2$ matrices. The symmetry conditions on the modified covariance matrix ($\tilde{\boldsymbol{\sigma}}$) in eq. (\ref{Sigmatilde}) are assumed to be preserved for the basis matrices ($\tilde{\boldsymbol{\sigma}}_{\alpha}$) as well and the solutions are computed based on it. Together with these conditions and eq. (\ref{CT_Lyapunov_w_base}), we get the following matrix relations
\begin{gather}
\boldsymbol{G}_{\alpha}^{T} = \boldsymbol{G}_{\alpha},  \label{G_matrix} \\
\boldsymbol{F}_{\alpha}^{T} = \boldsymbol{KJ}_{\alpha}/m,    \\
(\boldsymbol{KB}_{\alpha})^{T} = \boldsymbol{KB}_{\alpha},   \\
(\boldsymbol{KF}_{\alpha})^{T} = -\boldsymbol{KF}_{\alpha},    \\
\boldsymbol{KB}_{\alpha}+\boldsymbol{\Gamma F}_{\alpha} = \boldsymbol{G}_{\alpha}\boldsymbol{K},   \\
m\boldsymbol{F}_{\alpha}+m\boldsymbol{F}_{\alpha}^{T}+\boldsymbol{\Gamma G}_{\alpha}+\boldsymbol{G}_{\alpha} \boldsymbol{\Gamma} = 2\boldsymbol{E}_{\alpha}, \label{E_matrix}
\end{gather}
where $\boldsymbol{E}_{\alpha} =  \begin{bmatrix}
\delta_{\alpha L} & 0\\
0 & \delta_{\alpha R}
\end{bmatrix}$. The residual matrix may be expressed as
\begin{equation}
\tilde{\boldsymbol{\sigma}}^{r} = \gamma_{L}(T_{L}-T_{e})\tilde{\boldsymbol{\sigma}}_{L} + \gamma_{R}(T_{R}-T_{e})\tilde{\boldsymbol{\sigma}}_{R}.
\label{Sigma_r}
\end{equation}
By taking the trace of $\tilde{\boldsymbol{\sigma}}^{r}$ in the above and using the traceless property, we can obtain the effective temperature 
as $T_{e} = C_{L}T_{L}+C_{R}T_{R}$, where 
\begin{equation}
C_{\alpha} = \frac{\gamma_{\alpha}\text{Tr}\tilde{\boldsymbol{\sigma}}_{\alpha}}{\gamma_{L}\text{Tr}\tilde{\boldsymbol{\sigma}}_{L}+\gamma_{R}\text{Tr}\tilde{\boldsymbol{\sigma}}_{R}}.
\label{Coeff_Sigmar}
\end{equation}
The steady state distribution is $P(\boldsymbol{z}) = \exp^{-(H+\Delta H)/T_{e}}/\mathcal{Z}$ where the Hamiltonian ($H$)  is
\begin{eqnarray}
H = \frac{1}{2}m\boldsymbol{v}^{T}\boldsymbol{v}+\frac{1}{2}\boldsymbol{x}^{T}\boldsymbol{K}\boldsymbol{x} = \frac{1}{2}\boldsymbol{z}^{T}\tilde{\boldsymbol{K}}\boldsymbol{z},
\label{Hamiltonian_2B}
\end{eqnarray}
and the additional term ($\Delta H$) is
\begin{eqnarray}
\Delta H = \frac{1}{2}\boldsymbol{z}^{T}\tilde{\boldsymbol{K}}\left[(\boldsymbol{I}_{4}+\tilde{\boldsymbol{\sigma}}^{r}/T_{e})^{-1}-\boldsymbol{I}_{4} \right]\boldsymbol{z}.
\label{dH_2B}
\end{eqnarray}
We can solve for $\tilde{\boldsymbol{\sigma}}_{\alpha}$ using eqs. (\ref{G_matrix}-\ref{E_matrix}). Then, we get $\tilde{\boldsymbol{\sigma}}^{r}$ from eq. (\ref{Sigma_r}) .  Thus, if we know $\tilde{\boldsymbol{\sigma}}_{\alpha}$, we can determine $\tilde{\boldsymbol{\sigma}}^{r}$, $C_{\alpha}$ and $T_{e}$. The coefficients come out as
\begin{eqnarray}
C_{L} = \frac{2\kappa^{2}\gamma_{L}(\gamma_{L}+\gamma_{R})+\gamma_{L}\gamma_{R}\Omega}{2\kappa^{2}(\gamma_{L}+\gamma_{R})^{2}+2\gamma_{L}\gamma_{R}\Omega}, \nonumber \\
C_{R} = \frac{2\kappa^{2}\gamma_{R}(\gamma_{L}+\gamma_{R})+\gamma_{L}\gamma_{R}\Omega}{2\kappa^{2}(\gamma_{L}+\gamma_{R})^{2}+2\gamma_{L}\gamma_{R}\Omega},
\label{Coeff_Calc}
\end{eqnarray}
 where $\Omega = (k_{1}-k_{2})^{2}+(\gamma_{L}+\gamma_{R})(k_{1}\gamma_{R}+k_{2}\gamma_{L})/m$. 

In general, the weight factors ($C_{\alpha}$) are unequal to each other. If we assume symmetrical friction 
relative to both baths ($\gamma_{L}=\gamma_{R}$), then 
$C_{L}=C_{R}=1/2$ and $T_{e}=(T_{L}+T_{R})/2$.

\bibliography{Paper_BeadSpring_NEQ.bib}

\end{document}